\documentclass[12pt,nofootinbib]{revtex4-1}
\usepackage{amsmath,amssymb}
\usepackage{slashed,verbatim,graphicx}
\usepackage{color}

\newcommand\vev[1]{\langle #1\rangle}
\newcommand\ket[1]{| #1\rangle}
\newcommand\bra[1]{\langle #1|}

\newcommand{\beq}{\begin{equation}}
\newcommand{\beqs}{\begin{equation*}}
\newcommand{\eeq}{\end{equation}}
\newcommand{\eeqs}{\end{equation*}}
\newcommand{\CO}{  {\cal O}  }

\begin{document}
\setlength{\unitlength}{1mm}
%%%%%%%%%%%%%%%%%%%%%%%%%%%%%%%%%%%%%%%%%%%%%%%%%%%%%%%%%%%%%%%%
\title{Logarithmic enhancements  in conformal perturbation  theory and their real time interpretation.}

\author{David Berenstein, Alexandra Miller}
\affiliation { Department of Physics, University of California at Santa Barbara, CA 93106}

\begin{abstract} 
We study various corrections of correlation functions  to leading order in conformal perturbation theory, both on the cylinder and on the plane. Many problems on the cylinder are mathematically equivalent to those in the plane if we give the perturbations a position dependent scaling profile. The integrals to be done are then similar to the study of correlation functions with one additional insertion at the center of the profile. We will be primarily interested in the divergence structure of these corrections when computed in dimensional regularization.
 In particular, we show that the logarithmic divergences (enhancements) that show up in the plane under these circumstances can be understood in terms of resonant behavior in time dependent perturbation theory, for a transition between states that is induced by an oscillatory perturbation on the cylinder.
 \end{abstract}

\maketitle

%%%%%%%%%%%%%%%%%%%%%%%%%%%%%%%%%%%%%%%%%%%%%%%%%%%%%%%%%%%%%%%%
\section{Introduction }
\label{S:Introduction}

Since the advent of the AdS/CFT correspondence \cite{Maldacena:1997re}, the study of conformal field theories \footnote{This includes deformations away from the conformal fixed point.} has advanced substantially, as one can solve difficult problems in the dual gravity theory instead. 
Many of the results found this way are not particular to gravitational theories: they are universal in conformal field theory. For example, in the study of quenches, one can find the anomalous scaling of various dynamical expectation values \cite{Buchel:2013gba, Das:2014hqa}. This behavior can be understood from conformal perturbation theory, as shown in \cite{Berenstein:2014cia}.
In particular, in this last work, it was argued that many of the problems can be handled by the use of dimensional regularization on the cylinder, where one leaves the operator dimensions fixed, but integrates the angular variables in arbitrary number of spacetime dimensions.

In this paper, we study corrections to various correlation functions in conformal perturbation theory, using dimensional regularization techniques.  These results will be universal and can be applied to the gravitational theory via the AdS/CFT duality.  Specifically, we will analyze the divergence structure of these corrections and provide a novel interpretation of their origin.  

The general problem we are studying is that of a theory which has been deformed away from a conformal fixed point by a scalar operator $\lambda\int d^dx f (x) \CO_D(x)$ (the subscript $D$ standing for deformation).  We consider relevant, marginal, and irrelevant operators with dimension $h_D$ and work in arbitrary number of spacetime dimensions $d$.  Notice, we further allow the deformation to be spacetime dependent, including the function $f(x)$ and taking the constant $\lambda$ to be our small parameter.  Correlation functions in the new theory take the form
\begin{equation}
\left\langle \CO_1(x_1) \dots \CO_N(x_N) \right\rangle_{\lambda} = \left\langle \CO_1(x_1) \dots \CO_N(x_N) e^{\lambda\int d^dx f (x) \CO_D(x)}\right\rangle_{\rm{CFT}}
\end{equation}
where the correlators on the left hand side indicate the path integral in perturbed theory, while those on the right are in the conformal theory.  One can expand the right hand side to find
\begin{eqnarray}
\left\langle \CO_1(x_1) \dots \CO_N(x_N) \right\rangle_{\lambda} &=& \left\langle \CO_1(x_1) \dots \CO_N(x_N) \right\rangle_{\rm{CFT}} \\
&&+ \lambda\int d^dx f (x)  \left\langle \CO_1(x_1) \dots \CO_N(x_N) \CO_D(x)\right\rangle_{\rm{CFT}} + O(\lambda^2)
\end{eqnarray}

We assume that the conformal field theory in question has a  known set of operator dimensions (a spectrum of conformal representations), and known three point functions.  One then simply includes, for instance, the known form of the three point function
\begin{equation}
\vev{\CO_1(z) \CO_2(w) \CO_3(x)}_{\rm{CFT}} = \frac{f_{123}}{|z-w|^{h_1+h_2-h_3}|x-z|^{h_1+h_3-h_2} |x-w|^{h_2+h_3-h_1} }
\end{equation}
and only needs to integrate to find the desired corrections.  

However, these integrals can be tricky and are divergent.  Consider, for instance, the correction to the one point function when the theory is perturbed by a constant deformation
\begin{equation}
\vev{\CO_D(0)}_\lambda = \lambda\int \frac{d^d x}{|x|^{2h_D}} + O(\lambda^2)
\end{equation}
where we have used the known form of the two point function in a CFT.  This expression diverges either at the origin (a UV divergence) or at infinity (an IR divergence), or both if $h_D=d/2$.  Transforming the theory to the cylinder can provide an infrared regulator (the size of the cylinder).  The answer will then be IR finite, but there still might be UV divergences. This is what was studied by the present authors in \cite{Berenstein:2014cia}. In that paper, it was shown how to remove the divergences by using a modified version of dimensional regularization, where $d$ is varied, but the operator dimensions are fixed. This was shown to be very similar to keeping $d$ fixed and varying the dimensions of the operators: the results were expressed in terms of gamma functions of linear combinations of $d, h_D$. Logarithmic divergences occurred at special values associated with the pole structure of the gamma functions. This was also shown to be equal to the solution of the problem in the gauge/gravity duality.

Let's explicitly see how this IR divergence was tamed.  To transform the operators from the plane to the cylinder, one must introduce powers of the Weyl rescaling.  That is,
\begin{equation}
\left\langle \CO_1 \dots \CO_N \right\rangle_{\rm{cyl}} = |x_1|^{\Delta_1} \dots |x_N|^{\Delta_N} \left\langle \CO_1 \dots \CO_N \right\rangle_{\rm{plane}}
\end{equation}
It was this additional factor that helped with the convergence of the integral.  Interestingly, one can alternatively think of this factor of as having come from having made the deformation spacetime dependent, with a source that scales as $f(x) = |x|^{h_D-d}$
\begin{equation}
\vev{\CO(z)}_\lambda = \lambda\int d^dx f(x) \frac {1}{|x-z|^{2h_D}} +O(\lambda^2) = \lambda\int d^dx\frac 1{|x|^{d-h_D}} \frac {1}{|x-z|^{2h_D}} +O(\lambda^2)
\end{equation}
In this case, the infrared cutoff is provided by the fact that the profile of $f(x)$ dies sufficiently fast at infinity. (Note the additional factor of $|x|^d$ comes from the change of measure associated to the Weyl rescaling transformation.)

In this work, we will consider various functions that make these integrals more convergent and will discuss the physics of the divergences that remain.  In particular, we will see that the correlators have logarithmic enhancements and will show that some of the singularities that appear in the process of evaluating the integrals with dimensional regularization techniques can be understood in terms of secular (resonant) behavior in time dependent perturbation theory on the cylinder.

The paper is organized as follows: We first consider three physically interesting master integrals and analyze the divergences in the resulting expressions.  Next, we study the special case of marginal deformations.  And finally, we relate our computations to time dependent perturbation theory on the cylinder and see how the singularities can be interpreted as secular resonances that arise in time dependent perturbation theory. The poles and logarithmic enhancements in the plane computations end up reflecting the secular behavior for resonant perturbations in the cylinder.

\section{Dimensional regularization master integrals}

There are three integrals we will consider explicitly in this work.  They are all generically of the form
\begin{equation}
I= \int \frac{d^d x \, f(x)} {|x-x_1|^{\alpha_1}\dots |x-x_N|^{\alpha_N}}
\end{equation}
There is a vast array of literature on these Feynman integrals. For a general reference, we suggest \cite{Smirnov:2004ym}.

  And, in fact, each of our computations have been performed in some form elsewhere.  We include the details for completeness and so that one may track the divergence structure throughout the computation.  It will be this structure that we are ultimately concerned with.

As is standard practice in evaluating these types of integrals, we will find it to be very useful to introduce Schwinger parameters, given by
\begin{equation}
\frac 1{|B|^ {2 a} } = \frac{1}{\Gamma(a)}\int^\infty_0 dt t ^{a-1} \exp(- t |B|^2)
\end{equation}

\subsection{Fourier transform of the two point function}

The first integral we consider is
\begin{equation}
I_{\Delta}[\vec k; z] =\int d^d x |z-x|^{-2 \Delta} \exp(i k x)  
\end{equation}
Like in the previous work \cite{Berenstein:2014cia}, this gives the first order correction to the one point function in the presence of a deformation.  However, rather than regulating by transforming to the cylinder, here we introduce  a source that is oscillating in position space $f(x) \simeq \exp(i k x)$. These deformations can be studied in a dual gravitational theory, where they produce a lattice that breaks translation invariance \footnote{One can also do this by adding random disorder \cite{Hartnoll:2008hs}, which we will not study. } (see for example \cite{Horowitz:2012ky,Horowitz:2012gs}). 
Here, the IR regulator is provided by the scale of the oscillations.  The point is that for large distances, the integral is oscillatory and mostly cancels, removing the possible infrared divergence.  The small dimensionless parameter is $\lambda |k|^{\Delta -d}<<1$. 
This has already been computed in  other places \cite{Klebanov:1999tb, Coriano:2013jba,Bzowski:2013sza}.

We compute with varying $d$, keeping $\Delta$ fixed. One can see that integral is UV divergent if $2\Delta\geq d$, but otherwise should converge. This is because the large radius region is tamed by the oscillatory nature of the integral. In this sense, the momentum scale cuts off the possible infrared singularity.

 To perform the integral, we first write it in terms of a Schwinger parametrization
\begin{equation}
I_{\Delta}[\vec k; z]=\frac 1{\Gamma[\Delta]}\int d^d x' \int_0^\infty ds s^{\Delta -1} \exp(-s |x-z|^2) \exp(i k x) 
\end{equation}
The net result is that the integral becomes Gaussian in $x$, and can be done by the usual rules of dimensional regularization. Shifting first the integration variable from $x$ to $x'=x-z$,  we get
\begin{equation}
I_{\Delta}[\vec k; z]= \frac 1{\Gamma[\Delta]} \exp(i \vec k \vec z) \int d^d x' \int_0^\infty ds s^{\Delta -1} \exp(-s |x'|^2) \exp(i \vec k \vec x') 
\end{equation}
and then complete the square to find
\begin{equation}
I_{\Delta}[\vec k; z]= \frac 1{\Gamma[\Delta]} \exp(i \vec k \vec z) \int d^d x' \int_0^\infty ds s^{\Delta -1} \exp(-s |x'|^2) \exp(- |k|^2 / 4 s ) 
\end{equation}
The Gaussian integral is trivially done in $d$ dimensions, giving us
\begin{equation}
I_{\Delta}[\vec k; z]= \frac {\sqrt \pi^d}{\Gamma[\Delta]} \exp(i \vec k \vec z) \int_0^\infty ds s^{\Delta -1} s^{-d/2}  \exp(- |k|^2 / 4 s ) 
\end{equation}
Finally, we can change variables to $\tilde s = 1/s$, so that 
\begin{equation}
I_{\Delta}[\vec k; z]= \frac {\sqrt \pi^d}{\Gamma[\Delta]} \exp(i \vec k \vec z)  \int_0^\infty d\tilde s  \tilde s^{-\Delta -1} \tilde s^{d/2}  \exp(- \tilde s |k|^2 / 4 ) 
\end{equation}
which we immediately recognize as a gamma function. The final answer is 
\begin{equation}
I_{\Delta}[\vec k; z]= \frac { \Gamma[d/2-\Delta]}{\Gamma[\Delta]}\sqrt \pi^d \exp(i \vec k \vec z)  \left(|k|^2 / 4\right)^{\Delta-d/2}\label{eq:Fou1}
\end{equation}
Generically, the  UV divergences for $\Delta>d/2$ have been removed by analytic continuation. The integral is always UV convergent for large enough $d$, if $\Delta$ is kept fixed. This defines a function of $d,\Delta$ that can be continued to values where the naive integral has a UV divergence. 
  
Notice that there is a singularity whenever $\Delta -d/2$ is a non-negative integer. These arise as poles in the gamma function. 
The singularity at $\Delta= d/2$ is exactly a logarithmic divergence. For the other cases, the singularity is a subleading logarithmic divergence.
 To get a finite answer in those cases, we need to add a counterterm. The counterterm is a polynomial in $k^2$, multiplied by $1/\epsilon$, where $\epsilon = d-d_0$ is the small parameter that deforms the dimension $d$ away from the dimension $d_0$ of interest. Because it is polynomial in $k^2$, it is  local. This is a contact term. We write the full expression as follows
 \begin{equation}
 \vev{\CO(x) \CO(z)} \simeq |z-x|^{-2\Delta} + b_{CT} \square^{\Delta-d/2} \delta^d(x-z)
 \end{equation}
The full final answer needs to be expanded in Taylor series in $d-d_0$. This produces an extra logarithm from 
\begin{equation}
(|k|^2)^{d_0/2-d/2}\simeq 1- \frac{\epsilon}{2} \log(k^2/\mu^2)
\end{equation}
Combined with the pole in the gamma function we get an enhancement of the answer by a logarithm, where we have introduced a renormalization group scale $\mu^2$  for dimensional  reasons. The $\mu^2$ lets us shift the finite part of the counterterm to be whatever we wish it to be.

\subsection{First order correction to the two point function}

Another interesting profile is to consider a different scaling function as follows
\begin{equation}
f(x) \simeq 1/|x|^ {\alpha}
\end{equation}
where we can choose $\alpha$ to be real or complex. If we want $f(x)$ to be real, we can also take the real part of the expression. 
This leads to the same integral that appears when considering a correction to the two point function of two different primary operators $\CO_1(z), \CO_2(w)$ in conformal perturbation theory. In this case, the infrared regulator is provided by the distance between the operators.

If we consider two such primary operators $\CO_1$, $\CO_2$ of dimensions $h_1,h_2$ and a perturbation of the field theory by a scalar operator $\CO_D(x)$, then the two point function Green function  for the operators is 
\begin{equation}
\vev{\CO_1(z) \CO_2(w)}_{\lambda}= \frac{\delta_{h_1, h_2}}{|z-w|^{h_1+h_2}}+ \lambda \int d^d x \vev{\CO_1(z) \CO_2(w) \CO_D(x)}_{\rm{CFT}}+\dots
\end{equation}
where $\CO_D(x)$ is the operator that perturbs away from the conformal fixed point. The Kronecker delta appearing in the expression can generically depend on spin labels of the operators, and the direction
vector between $z,w$. We will be interested in the simplest setting where both $\CO_1,\CO_2$ are  primary scalar operators (and so is $\CO_D$, in order not to break rotational symmetry).
 If $h_1\neq h_2$, the first term vanishes, as the two operators  then have a vanishing two point function in the conformal field theory.
The fact that the right hand side does not generically vanish beyond the leading order in conformal perturbation theory will be referred to as operator mixing.

If we use the known form of the three point function in a CFT, we see that the integral we wish to perform is
\begin{equation}
  I[z,w, h_D,d, h_1, h_2] = \int d^d x \frac 1{|x-z|^{h_1+h_D-h_2} |x-w|^{h_2+h_D-h_1}}\label{eq:master2}
\end{equation}
As can be seen, the integral diverges in the infrared if in the asymptotic $|x|\to \infty$ region we have that 
\begin{equation}
 \int_{x_0}^\infty d^d x \frac 1{|x|^{2 h_D}}
\end{equation}
is divergent. The integral is infrared convergent  if $2 h_D\geq d$.

Similarly, the integral is UV divergent for $x\simeq z$ if $h_1+h_D-h_2\geq d$. The same is true near $x\simeq w$ if $h_2-h_1+h_D\geq d$, and these follow from keeping the most singular terms near each one 
of the insertions of the operators.

Adding these two, we find that there is always a UV divergence if  $h_D\geq d$. That is, if the operator that performs the deformation $\CO_D(x)$ is marginal or irrelevant. The divergences then need to be regulated before getting the correct (renormalized) physical answer.

A standard procedure in the literature is to perform a geometric cutoff: do the integrals until we are within a distance $\delta x< \Lambda^{-1}$ (see for example \cite{kadanoff1979multicritical,Burrington:2012yq}). This is problematic at higher orders. These integrals can also be handled via dimensional regularization, which is the procedure we will follow. Here we keep $h_1, h_2, h_D$ fixed, and evaluate the integrals  for a variable complex $d$. The integral is defined in the non-convergent region by  analytically continuing in $d$ past the singularities.

We will now perform the integral.  Again, we start with Schwinger parameterization, which is valid for $a\geq 0$. With this, we find
\begin{equation}
 I[z,w, h_D,d, h_1, h_2] = \frac{\int d^d x \int_0^\infty d t_1 \int_0^\infty d t_2  t_1 ^{\left(\frac{\Delta h+h_D}{2}\right)-1} t_2^{\left(\frac{h_D-\Delta h }{2}\right)-1} }{\Gamma\left(\frac{h_1+h_D-h_2}{2}\right)\Gamma\left(\frac{h_2+h_D-h_1}{2}\right)} 
 \exp(-t_1 |x-z|^2-t_2 |x-w|^2)
\end{equation}
where we have introduced $\Delta h = h_1-h_2$. 
The Schwinger parametrization is allowed as long as both 
$h_D\pm \Delta h\geq 0$ and it is defined  for other values of these quantities  by analytic continuation, contingent on the $\Gamma$ function being evaluated at  a non-singular value (the singularities occur when 
$2(|\Delta h|-h_D)$ is a non-negative integer). As we see, the net result is that integral over $x$ again becomes Gaussian. 

It is convenient to change variables to $t_1= t y, t_2=t (1-y)$, so that $dt_1 d t_2= t dt dy$, so that $t= t_1+t_2$ as is usually done with Feynman parameters. We get then that
\begin{eqnarray}
 I[z,w, h_D,d, h_1, h_2] &=& \frac{\int d^d x \int_0^\infty t^{h_D-1} d t \int_0^1  d y  y ^{\left(\frac{\Delta h+h_D}{2}\right)-1} (1-y)^{\left(\frac{h_D-\Delta h }{2}\right)-1} }{\Gamma\left(\frac{h_1+h_D-h_2}{2}\right)\Gamma\left(\frac{h_2+h_D-h_1}{2}\right)} \nonumber\\
&& \times \exp(-t y |x-z|^2-t (1-y) |x-w|^2)\label{eq:step_inte}
\end{eqnarray}
We now complete the square to do the Gaussian integral over $x$, to get that 
\begin{eqnarray}
 I[z,w, h_D,d, h_1, h_2]&=&\frac{ \int_0^\infty t^{h_D-1} d t \int_0^1  d y  y ^{\left(\frac{\Delta h+h_D}{2}\right)-1} (1-y)^{\left(\frac{h_D-\Delta h }{2}\right)-1} }{\Gamma\left(\frac{h_1+h_D-h_2}{2}\right)\Gamma\left(\frac{h_2+h_D-h_1}{2}\right)}\nonumber \\
&& \times\frac{ \sqrt{\pi}^d  \exp(-t y (1-y) |z-w|^2)}{ t^{d/2}}\label{eq:step_gauss}
\end{eqnarray}
The Gaussian integral has been evaluated in an arbitrary number of (complex) dimensions $d$ by analytic continuation from positive integer dimension, as is standard in dimensional regularization. 

What we need to do now is understand the region of $t,y$ plane that picks up the singularities corresponding to $x\simeq z$ or $x\simeq w$ from passing from the equation \eqref{eq:step_inte} to equation \eqref{eq:step_gauss}. Obviously, the Gaussian is convergent as long as $t>0$, so all the $UV$ singularities are related to the region near $t\simeq 0$. In particular, $t_1$ small is the UV region of the singularity at $z$, and $t_2$ near zero is the UV region near the singularity at $w$.

Upon integration in $t$, we get that
\begin{equation}
 I[z,w, h_D,d, h_1, h_2]= \frac{ \int_0^1  d y  y ^{\left(\frac{\Delta h+h_D}{2}\right)-1} (1-y)^{\left(\frac{h_D-\Delta h }{2}\right)-1} }{\Gamma\left(\frac{h_1+h_D-h_2}{2}\right)\Gamma\left(\frac{h_2+h_D-h_1}{2}\right)}  \times\frac{ \sqrt{\pi}^d  \Gamma[h_D-d/2] }{\left[ y (1-y) |z-w|^2)\right]^{(h_D-d/2)}}
\end{equation}
Pulling out the constants that do not need to be integrated further, we find
\begin{eqnarray}
 I[z,w, h_D,d, h_1, h_2]&=& \frac{ \sqrt{\pi}^d  \Gamma[h_D-d/2] }{\Gamma\left(\frac{h_1+h_D-h_2}{2}\right)\Gamma\left(\frac{h_2+h_D-h_1}{2}\right)|z-w|^{ 2h_D-d}}\nonumber\\
 &&\times \int_0^1  d y  y ^{\left(\frac{d+\Delta h-h_D}{2}\right)-1} (1-y)^{\left(\frac{d-\Delta h-h_D }{2}\right)-1}
\end{eqnarray}
The dependence on $|z-w|$ could have been guessed by dimensional analysis. This step also leads to a Gamma function with a singularity at $h_D= d/2$. This is the infrared singularity that appears in the integral for large $x$. 

We now get, upon performing the integral over the last remaining variable, that the full answer is 
\begin{equation}
 I[z,w, h_D,d, h_1, h_2]= \frac{ \sqrt{\pi}^d  \Gamma[h_D-d/2] }{|z-w|^{ 2h_D-d}\Gamma\left(\frac{h_1+h_D-h_2}{2}\right)\Gamma\left(\frac{h_2+h_D-h_1}{2}\right)}\frac{\Gamma\left(\frac{d+\Delta h - h_D}2\right)\Gamma\left(\frac{d-\Delta h - h_D}2\right)}{\Gamma(d-h_D)}\label{eq:final_ans}
\end{equation}
This is the correct answer as long as the two gamma functions in the numerator have a positive argument. This in particular requires that $d> h_D$. In other regions, the result is defined by analytic continuation in $d$, first evaluated at large $d$, and then we bring down $d$ to the physical dimension of interest.

This final answer is very similar to the answers one gets from regular dimensional regularization of Feynman diagrams in field theory. Indeed, the integrals that have been done are of the same type.
Here the improvements in the answer are obtained by taking large $d$ first. This is because we are not allowing the dimension of the operators to change as we change $d$. Thus, marginal operators become relevant as we take $d$ large and keep the dimension of the deformation $h_D$ fixed. 

Singularities in the final answer occur  when $h_D\pm \Delta h -d$ is an even non-negative integer. These appear as poles of the gamma function. There is also a pole in the denominator that occurs if $h_D-d$ is a non-negative  integer. In all other cases, where there are no singularities, we have obtained a finite answer. This is the dimensionally regularized answer for the correction to the two point function.

In terms of the exponent variables (which we call $\alpha,\beta$), this integral is given by 
 \begin{eqnarray}
I[z-w, \alpha, \beta, d]&=&\int d^d y \frac 1{|x-w|^{2\alpha} |x-z|^{2\beta}}
\\
&=&\frac{\pi ^{d/2} \Gamma \left(\frac{d-\alpha }{2}\right) \Gamma \left(\frac{d-\beta
   }{2}\right) \Gamma \left(\frac{1}{2} (-d+\alpha +\beta )\right) (z-w)^{-\alpha -\beta
   +d}}{\Gamma \left(\frac{\alpha }{2}\right) \Gamma \left(\frac{\beta }{2}\right) \Gamma
   \left(d-\frac{\alpha }{2}-\frac{\beta }{2}\right)}\label{eq:alphabeta}
\end{eqnarray}

\subsection{First order correction to the three point function}

To get a correction to a three point function, we usually need to integrate a four point function of the form
\begin{equation}
\vev{\CO_1(\omega_1) \CO_2(\omega_2) \CO_3(\omega_3)}_\lambda =\int d^d y \vev{\CO_1(\omega_1) \CO_2(\omega_2) \CO_3(\omega_3) \CO_D(y)} 
\end{equation}
in the vast majority of the cases, such four point functions are not known exactly.

If one does perturbation theory with operators that are polynomials in a scalar field, or in the example of string scattering amplitudes in flat space, the integrals one needs to perform reduce to a finite number of integrals of the type
\begin{equation}
I[\overrightarrow{\omega}, \alpha_1 , \alpha_2 , \alpha_3, d]= \int d^d x \frac{1}{|x-\omega_1|^{2\alpha_1} |x-\omega_2|^{2\alpha_2} |x-\omega_3|^{2\alpha_3}}
\end{equation}
The precise details of the integral evaluation can be found in the appendix \ref{app:a}. The techniques are similar to the ones used before, but in general the answer is not particulalry simple.

It turns out that a similar integral is also obtained if we are working on a correction of a two point function where we have added some position dependence to the deformation. We will study this particular case in detail later on.

\section{Marginal deformations}

It is interesting to analyze the special case of marginal deformations, where $h_D=d$ in equation \eqref{eq:final_ans}. In that case, we find the following: there is always a pole in the denominator at $h_D=d$. This means that unless the numerators are singular, the answer actually vanishes. 
This is expected from the usual rules of unitary conformal field theories. Two point functions of primary operators of different dimensions should vanish at a conformal fixed point.

For a singularity in the numerator to occur we require that $\Delta h$ is an even integer.  There are two cases of interest. When $\Delta h\neq 0$ and the special case where $\Delta h=0$. Let us first analyze the case where $\Delta h\neq 0$, but still an even integer. In that case we find that  the answer is equal to
\begin{equation}
 I[z,w, h_D,d, h_1, h_2]= \frac{ \sqrt{\pi}^d  \Gamma[h_D-d/2] }{|z-w|^{ 2h_D-d}\Gamma\left(\frac{\Delta h+h_D}{2}\right)\Gamma\left(\frac{h_D-\Delta h}{2}\right)}\frac{\Gamma\left(\frac{d+\Delta h - h_D}2\right)\Gamma\left(\frac{d-\Delta h - h_D}2\right)}{\Gamma(d-h_D)}\label{eq:final_ans}
\end{equation}
If we keep $h_D$ fixed and analytically continue in $d$ we find that only one of the two numerators can become singular. The answer in the limit is of the form 
\begin{equation}
\lim _{\epsilon\to 0}\Gamma[\epsilon/2-m]/\Gamma[\epsilon] \times \hbox{finite} =\hbox{finite}\end{equation}
and this suggests that there could be finite mixing.

However, because in this case we have that $\Delta h = 2m $, the operators $\CO_1(x)$ and $\square^m \CO_2(x)$ have the same dimension. It is easy to see that we can modify $\CO_1$ with a finite
counterterm $\CO_1-c \square^m \CO_2$ that removes the mixing. In this sense, this is no different that $\Delta h\neq 0$ and we see the absence of mixing. Such a term can be interpreted as a contact term in the OPE. It is also in this case, when the dimension $h_D$ is even, that there can also be an extra pole in the denominator that arises from the Schwinger parametrization of the denominators. This can produce a double pole in the denominator and makes the end result vanish.

Now, only the special case where $\Delta h =0$ and the operator is marginal remains to be studied.
The limit looks like
\begin{equation}
\lim_{\epsilon\to 0} \Gamma[\epsilon/2] \Gamma[\epsilon/2] / \Gamma[\epsilon]\simeq 4/\epsilon+\hbox{finite}
\end{equation} 
In this case we produce a universal logarithm. This is a correction to the anomalous dimension of the operator $\CO_1$. It depends only on the OPE coefficient
\begin{equation}
h_D(y)\CO_1(x)\simeq f_{11D}|x-y|^{-h_D} \CO_1(x)+\dots
\end{equation}
 but not on the dimension of the operator $h_1$. In that sense, the integral we have to perform is always universal.

\section{Position and time dependent perturbations}

As described in the introduction, the second master integral that we evaluated in the previous section can also be interpreted in terms of a position dependent excitation on the cylinder. 
To see this, we convert the variables from the integral to the natural ones on the cylinder.

To first order, the correction to the one point function on the cylinder in the presence of a deformation with an $f(x)=e^{i\omega\tau}$ factor is given by
\begin{equation}
\vev{\CO_D(\Omega',\tau')}_{cyl} = \lambda \int_{cyl} d^{d-1} \Omega d\tau \exp^{i \omega \tau}\vev{\CO_D(\Omega, \tau) \CO_D(\Omega, \tau')}_{cyl}
\end{equation}
where $\tau$ is the natural euclidean time coordinate on the cylinder and $\omega$ is a complex variable. When $\omega$ is real we have a bounded and oscillating perturbation of the conformal field theory on the cylinder.

Using the fact 
that in radial quantization we have $\tau \simeq \log r$, and that to convert to the punctured plane we need extra factors of $|r|^{-h_D}$ to be inserted, we get that 
\begin{eqnarray}
\int_{cyl} d^{d-1} \Omega d\tau  \exp^{i \omega \tau}\vev{\CO(\Omega, \tau) \CO(\Omega, \tau')}_{cyl} &=& \int d^d x |x|^{i\omega+h_D-d} |y|^{h_D} \vev{\CO(x) \CO(y)}\\
&=&\int d^d x |x|^{i\omega+h_D-d} |y|^{h_D}\frac{1}{|x-y|^{2h_D}}\label{eq:w-dep}
\end{eqnarray}
This can be written as
\begin{equation}
\int_{cyl} d^{d-1} \Omega d\tau  \exp^{i \omega \tau}\vev{\CO(\Omega, \tau) \CO(\Omega, \tau')}_{cyl} = I(|y|,\alpha, \beta, d) |y|^{h_D}
 \end{equation}
 provided we identify
 \begin{equation}
\alpha =- h_D - i \omega+d, \beta = 2 h_D
 \end{equation}
 in the master integral \eqref{eq:alphabeta}.
 The result is then given by
 \begin{equation}
\frac{\pi ^{d/2} \Gamma \left(\frac{1}{2} \left(d-2 h_D\right)\right) \Gamma
   \left(\frac{1}{2} \left(h_D-i \omega \right)\right) \Gamma \left(\frac{1}{2} \left(i
   \omega +h_D\right)\right) z^{h_D+i \omega }}{\Gamma \left(h_D\right) \Gamma
   \left(\frac{1}{2} \left(d-i \omega -h_D\right)\right) \Gamma \left(\frac{1}{2}
   \left(d+i \omega -h_D\right)\right)}\label{eq:1ptcylw}
 \end{equation}
 
 This has the expected Euclidean time dependence. 
 Notice that this has singularities where
 \begin{equation}
 i \omega = \pm(h_D + 2 k)
 \end{equation} 
 and $k$ an integer. This is natural, as when we go from Euclidean to Lorentzian signature in a Wick rotation, we should make $\omega= i \tilde \omega$ imaginary in order to obtain a real frequency $\tilde\omega$. This then corresponds to driving the field theory on the cylinder with a time dependent source at frequency $\tilde \omega$.  At the values of $\pm\tilde \omega= h_D+2k$ we obtain resonances. These happen exactly at the energies of the spherically invariant excitations of the scalar $\CO$ on the cylinder (see \cite{Berenstein:2014cia} for a description of driving the conformal field theory in Hamiltonian mechanics).  

The exponent is also given by  
\begin{equation}
i\omega-h_D+h_D = i \omega
\end{equation}
as one expects from the translation properties of the integral in the cylinder coordinates. The way we deal with the resonance in this case is to analytically expand in $\tilde\omega$ at the required frequency and keep the first subleading term. This gives the usual secular growth of the resonance as $t \exp(i \tilde \omega  t)$.

This response can also be obtained from the AdS dual following similar steps to those found in \cite{Berenstein:2014cia}. We need to solve the differential equation for the radial coordinate
\begin{equation}
-\frac 1 {r^{d-1}} \partial_r((1+r^2) r^{d-1}\partial_r \phi(r)+m^2 f(r) +\tilde\omega^2/(1+r^2) \phi(r) =0
\end{equation}
This is in a coordinate system where
\begin{equation}
ds^2 \simeq -dt^2(1+r^2) + dr^2(1+r^2)^{-1} + r^2 d\Omega^2
\end{equation}

The problem of the asymptotic shape near $r\to \infty$ ends up  ends up being controlled by the asymptotic expansion of the hypergeometric function
\begin{equation}
\phi(r) \propto  \ _2F_1\left(\frac{ (d - h_D-  \tilde w)}{2}, \frac{(h_D- \tilde w)}2, \frac d2, -r^2\right)
\end{equation}
 with the usual relation between the mass in AdS and the dimension of the operator $m^2= \sqrt{h_D^2- d h_D}$. This produces the correct ratio of the $\Gamma$ functions.  The factors of $\pi$ etc, are explained in detail in \cite{Berenstein:2014cia}.

 We can also notice that from this result we can recover the Fourier transform of the two-point function in equation \eqref{eq:Fou1}. The idea is simple. If we take $\omega\to \infty$
 in equation \eqref{eq:1ptcylw}, we are driving the theory in the cylinder with a Euclidean time dependence that is oscillatory and that has a wavelength that is much smaller than the size of the cylinder. In this limit, we would expect that if we zoom in to the region where we have the operator, the infrared cutoff scale induced by the driving of the field theory is at a higher energy than the infrared cutoff provided by the geometry. This second cutoff should become invisible, up to "finite size" corrections. To take the limit, we use Stirlings approximation for the gamma functions that depend on $\omega$,  $\Gamma(\gamma) \simeq \exp( \gamma\log \gamma- \gamma)$.
 
 We find this way that in the limit the answer becomes
 \begin{equation}
\vev{\CO_D(\tau)}=\frac{\pi ^{d/2} \Gamma \left(\frac{1}{2} \left(d-2 h_D\right)\right)}{\Gamma \left(h_D\right)} \exp(i\omega\tau)\times F(\omega)
 \end{equation}
 where 
 \begin{equation}
F(\omega)= \exp\gamma_1\log \gamma_1-\gamma_1+\gamma_2\log(\gamma_2)-\gamma_2
-\gamma_3\log(\gamma_3)+\gamma_3-\gamma_4\log(\gamma_4)+\gamma_4) +O(1/\gamma)
 \end{equation}
where the $\gamma_i$ are the various $\omega$ dependent variables that appear as arguments in the Gamma functions. A straightforward evaluation shows us that
\begin{equation}
F(\omega)=\exp( (d-2 h_D)\log(2)+(2 h_D-d)\log(\omega)) = \left(\frac{\omega^2}{4}\right)^{h_D-d/2}
\end{equation}
and this matches the Fourier transform with all the factors of two and normalization factors on the nose.

We can go one step further. We can also consider the case where we compute a correction to a two point function in the presence of a position dependent perturbation with a radial profile. 
This is handled by the integral
\begin{equation}
\vev{\CO_1(x_1) \CO_2(x_2)} _{\omega}\simeq f_{12D} \int d^d x \frac{|x|^{i \omega+h_D-d}}{{|x-x_1|^{\Delta h+h_D}{|x-x_2|^{-\Delta h+h_D}}}} \label{eq:twopointf}
\end{equation}
where we have used the same convention for the profile as in equation \eqref{eq:w-dep}, so that it corresponds to an oscillating driving of the cylinder after the conformal rescaling that places the origin at the infinite past.
This gives rise to a more complicated integral. But, in the special case where either $x_1=0$ or $x_2=0$, it takes the same form as the simpler master integral we have already evaluated.

What is important for us, is that  this modifies the exponents in the usual three point function as follows
\begin{equation}
\Delta h+h_D\to \Delta h+h_D-i \omega-h_D+d= \Delta h - i \omega +d
\end{equation} 
keeping the other one,  $h_D-\Delta h $, fixed.

The $\omega$ dependent gamma factors end up being given by
\begin{equation}
\frac{\Gamma\left[\frac{h_D-i \omega}2\right]\Gamma\left[\frac{i \omega-\Delta h}2\right]}{\Gamma\left[\frac{d+\Delta h -i \omega}2\right]\Gamma\left[\frac{d-h_D +i \omega}2\right]}
\end{equation} 
when we choose $x_1$ to be at the origin. 

One set of poles in the numerator occur when $\tilde\omega =- i \omega = 2k-\Delta h $. This again can be interpreted as a resonance. After all, putting the operator $\CO_1$ at the origin puts the field theory in the vacuum of the representation of the conformal group associated to $\CO_1$ (the lowest weight state), which is spherically invariant. This is an immediate consequence of the conformal rescaling that takes the plane  to the cylinder.
This has energy $h_1$ relative to the usual vacuum. Because we are integrating with a spherically invariant perturbation profile, the states that are generated need to be spherically invariant. We can think of this as an amplitude in the cylinder Hamiltonian theory of the form
\begin{equation}
\int  d\tau \exp(i \tilde \omega \tau) \sum_k s_k \bra{\CO_2, 2k, \tau_2} \CO_D(\tau) \ket{\CO_1, 0,-\infty}
\end{equation}
where the ground state of the $\CO_1$ representation is converted into a sum of states in the $\CO_2$ representation by the action of the perturbation. The coefficients $s_k$
appear from how the operator $\CO_2$ at a particular radial time excites the individual states. This is explained in \cite{Berenstein:2014cia}.
Since $\ket{\CO_1,0,-\infty}$ is an eigenstate of the radial Hamiltonian, we can choose the initial time to be anywhere we want, and we can make that coincide with the lower end of the integration in the variable $\tau$.

Here, spherical symmetry of the initial state and the perturbation guarantees that the operator $\CO_2$ can only destroy spherically invariant states.  These can only have energies  $h_2+2k$ with $k$ an integer. 
In this sense, we can schematically write
\begin{equation}
\CO_D \simeq a_D^\dagger + f_{12D}\sum f_k a^\dagger_{2, k} a_{1,0}
\end{equation}
where the $f_k$ are determined by the conformal symmetry (this is related to the conformal block structure of the OPE between the representations $\CO_1,\CO_2, \CO_D$, it basically describes how descendant amplitudes are related to the primary amplitudes). 
The gamma function pole happens exactly at resonance for a transition between a state with energy $h_1$ and a state of energy $h_2+2k$. The resonance in this case is constructive interference between the perturbations at different times.
Again, in time dependent perturbation theory such resonances produce secular behavior (linear growth in time), which becomes a logarithm after passing from the cylinder to the plane.

Let us explain this. The usual formula for time dependent perturbation theory is 
\begin{equation}
\ket f = P \exp\left(\int i V_{int}(t)dt\right) \ket i
\end{equation}
where $V$ is in the interaction picture. This has a natural time dependence of $\exp(i (w_f-w_i )t)$ for a transition between states of energies labeled by their  frequencies $w_i, w_f$.
If we add to the problem an external time dependence at frequency $\omega$ we get  that the driving of the transition oscillates at a  shifted frequency $\omega-\Delta w$. This gives us a linearly growing transition in time if $\omega-\Delta w=0$, this is the secular term in perturbation theory. 
In our case the frequency of the final state is any of the spherically invariant descendants of $\CO_2$

 There is a second pole. This one does not depend on $\Delta h$. Instead, it corresponds to the topology for a transition of the form
 \begin{equation}
\bra{\CO_D, 2k, \tau}\CO_2(\tau_2)\ket{\CO_1, -\infty}
 \end{equation}
 where the secular behavior is produced because it has a time dependence that exactly cancels the time dependence of $\bra{\CO_D,2k,\tau}$ in the Schr\"odinger picture. This is why $\omega$ ends up with the other sign in the pole of the gamma function.
 
 Using the results of appendix \ref{app:a}, we can actually solve the full problem without restricting to putting one of the operators at the origin. This way we obtain the full correction to the  two point function on the cylinder. We will now show that many of the singularities in the final answer have the same interpretation. The idea is to do an integral of the form
 \begin{equation}
 I \simeq \int d^d x \frac{1}{|x-w_0|^{2a}|x-w_1|^{2b}|x-w_2|^{2c}}
 \end{equation}
 with $2 a= (d-h_D)-i \omega$, $2 b= h_D+\Delta h$, $2c = h_D-\Delta h$.
 The answer is then given by
 \begin{equation}
{\cal  N} \int_{0}^{\infty}{dt\left(\sqrt{t}\right)^{d/2-2}}\prod_{j=1}^{3}\Omega_{j}^{-a_{t}+d/2+a_{j}}K_{d/2+a_{j}-a_{t}}\left(\sqrt{2t} \Omega_j \right)
\end{equation}
where $a_t= a_1+a_2+a_3$ and ${\cal N}$ is a normalization factor given by 
\begin{equation}
{\cal N}= \frac{\pi^{d/2} 2^{a_t-d}}{\Gamma\left[a_{1}\right]\Gamma\left[a_{2}\right]\Gamma\left[a_{3}\right]\Gamma\left[d-a_{t}\right]}
\end{equation}
which is non-singular. The functions $K$ are modified Bessel functions. It is convenient to change variables to $x\propto \sqrt t$. 

There are two types of singularities that can show up. Some of them result from integration over the variable $x$, and others result from the normalization of the modified Bessel function, whose expansion is as follows
\begin{equation}
K_\nu(x) \simeq x^\nu 2^{-1-\nu} \Gamma[-\nu] (1+O(x^2))+2^{-1+\nu} x^{-\nu} \Gamma[\nu](1+O(x^2))
\end{equation}
The singularities all arise from the $x\simeq 0$ region. At large $x$, $K_\nu(x) \simeq \exp(-x)\times \hbox{power law}$, so the $x\to \infty$ limit of the  integral  is convergent.

 Singularities in the integral arise when $d/2-1\pm \nu_1\pm \nu_2\pm \nu_3$ is  a negative integer \cite{Bzowski:2015yxv}, where the $\nu$ are the labels of the Bessel function. But notice that 
 when we compute, we find that 
 \begin{eqnarray}
 a_2+a_3&=& h_D\\
 2 a_t&=& (d-h_D)-i \omega+ 2 h_D
 \\&=& d+h_D-i\omega\\
 d/2+a_1-a_t&=&d/2- h_D\\
 d/2+a_2-a_t &=& i \omega/2+ \Delta h/2 \\
 d/2+a_3-a_t&=& i \omega/2-\Delta h/2
 \end{eqnarray}
 so only two of the labels of the Bessel functions depend on $\omega$ and $\Delta h$, and not the third. When we take the combinations $\pm \nu_2\pm\nu_3$, either the dependence on $\Delta h$ or the dependence on $\omega$ cancels, so the  singularities in $\omega$ that arise from the integrals do not depend on $\Delta h$, which is what we are seeking to find.

Thus, the singularities we want to analyze must arise from the normalization factors of the modified Bessel functions. These are in the  $\Gamma$ factors. Poles will arise whenever
$\pm (i \omega/2\pm \Delta h/2)$ are integers. That is the same as writing
\begin{equation}
i\omega = \pm \Delta h \pm 2k
\end{equation}
Half of these singularities are transitions where a descendant of the first operator (in the initial state) is excited at a  resonant frequency  with descendants of the second operator.
The point is that the first operator not being at the origin produces a linear combination of the lowest energy state in the representation and it's descendants. Any one of which could be the one in resonance. The second set of singularities arises from reversing the order of the operators (thinking of $\CO_2$ as generating the initial state, rather than $\CO_1$). Both are generally required because in the Euclidean answer we can interchange the order of the operators without encountering a singularity. Again, spherical symmetry of the perturbation forces the state generated by $\CO_1$ to have the same angular quantum numbers as the state annihilated by $\CO_2$ (or viceversa). This is what produces a difference that is twice an integer, rather than just an integer. After all, states with the same angular momentum that are descendants of a single state, differ in their energies by twice an integer.

The upshot is that the poles in $\omega$ that depend on $\Delta h$  can always be interpreted in terms of resonant transitions.

  \acknowledgements
 
We woul like to thank K. Skenderis for conversations. Work supported in part by the U.S. Department of Energy under grant DE-SC0011702. 
D.B. and A.M. are very grateful to the Galileo Galilei Institute for their support where part of of this work took place.

 \appendix

 \section{Integral with three centers}\label{app:a}
 
We compute the integral given by

\begin{equation}
I[\overrightarrow{\omega}, \alpha_1 , \alpha_2 , \alpha_3, d]= \int d^d x \frac{1}{|x-\omega_1|^{2\alpha_1} |x-\omega_2|^{2\alpha_2} |x-\omega_3|^{2\alpha_3}}
\end{equation}
This integral was originally studied in \cite{Boos:1987bg}, where the answer is given in terms of an Appell Function, see also \cite{Coriano:2013jba}. We follow instead 
the treatment of \cite{Bzowski:2013sza}, which expresses the final answer in terms of an integral of modified Bessel functions.

We will introduce a Schwinger parameter for each term in the denominator.  This gives

\begin{equation}
=\int d^{d}x\int_{0}^{\infty}\frac{ds_{1}\,ds_{2}\,ds_{3}\,s_{1}^{\alpha_{1}-1}\,s_{2}^{\alpha_{2}-1}\,s_{3}^{\alpha_{3}-1}}{\Gamma\left[\alpha_{1}\right]\Gamma\left[\alpha_{2}\right]\Gamma\left[\alpha_{3}\right]}\,\exp\left(-s_{1}\left|x-\omega_{1}\right|^{2}-s_{2}\left|x-\omega_{2}\right|^{2}-s_{3}\left|x-\omega_{3}\right|^{2}\right)
\end{equation}

Completing the square in the exponential gives

\begin{eqnarray}
&=&\int d^{d}x^{\prime} \int_{0}^{\infty}\frac{ds_{1}\,ds_{2}\,ds_{3}\,s_{1}^{\alpha_{1}-1}\,s_{2}^{\alpha_{2}-1}\,s_{3}^{\alpha_{3}-1}}{\Gamma\left[\alpha_{1}\right]\Gamma\left[\alpha_{2}\right]\Gamma\left[\alpha_{3}\right]} \nonumber\\
&&\times \exp\left(-s_{t}\left[\left(x^{\prime}\right)^{2}-\frac{\left(s_{1}\omega_{1}+s_{2}\omega_{2}+s_{3}\omega_{3}\right)^{2}}{s_{t}^{2}}+\frac{\left(s_{1}\omega_{1}^{2}+s_{2}\omega_{2}^{2}+s_{3}\omega_{3}^{2}\right)}{s_{t}}\right]\right)
\end{eqnarray}

where $x^{\prime}=x-\frac{\left(s_{1}\omega_{1}+s_{2}\omega_{2}+s_{3}\omega_{3}\right)}{s_{t}}$ and $s_t = s_1 +s_2 +s_3$.  We can now easily perform the integral over $x^{\prime}$ because it is gaussian.   This all gives

\begin{equation}
I=\pi^{d/2}\int_{0}^{\infty}\frac{ds_{1}\,ds_{2}\,ds_{3}\,s_{1}^{\alpha_{1}-1}\,s_{2}^{\alpha_{2}-1}\,s_{3}^{\alpha_{3}-1}}{\Gamma\left[\alpha_{1}\right]\Gamma\left[\alpha_{2}\right]\Gamma\left[\alpha_{3}\right]s_{t}^{d/2}}\,\exp\left[\frac{\left(s_{1}\omega_{1}+s_{2}\omega_{2}+s_{3}\omega_{3}\right)^{2}}{s_{t}}-\left(s_{1}\omega_{1}^{2}+s_{2}\omega_{2}^{2}+s_{3}\omega_{3}^{2}\right)\right]
\end{equation}

We can simplify the exponential, leaving

\begin{equation}
I=\pi^{d/2}\int_{0}^{\infty}\frac{ds_{1}\,ds_{2}\,ds_{3}\,s_{1}^{\alpha_{1}-1}\,s_{2}^{\alpha_{2}-1}\,s_{3}^{\alpha_{3}-1}}{\Gamma\left[\alpha_{1}\right]\Gamma\left[\alpha_{2}\right]\Gamma\left[\alpha_{3}\right]s_{t}^{d/2}}\,\exp\left[ -\frac{s_{1}s_{2}\omega_{12}^{2}+s_{2}s_{3}\omega_{23}^{2}+s_{3}s_{1}\omega_{31}^{2}}{s_{t}} \right]
\end{equation}

Now introduce a change of variables 
\begin{equation}
s_{j}=\frac{V}{2v_{j}}=\frac{v_{1}v_{2}+v_{2}v_{3}+v_{3}v_{2}}{2v_{j}}
\end{equation}
These result from finding a change of variables such that $ s_1 s_2 s_3 /(s_t s_i) \propto v_i$, and the factor of two is the same convention as in the appendix in \cite{Bzowski:2013sza}.

The measure will change as $ds_{1}ds_{2}ds_{3}=\frac{V^{3}}{8v_{1}^{2}v_{2}^{2}v_{3}^{2}}dv_{1}dv_{2}dv_{3}$.  Our integral now takes the form

\begin{eqnarray}
=\pi^{d/2}\int_0^\infty\frac{V^3}{8 \Gamma^3 v_{1}^{2}v_{2}^{2}v_{3}^{2}} \frac{dv_{1}dv_{2}dv_{3}}{\left[\frac V2(v_1^{-1}+v_2^{-1}+v_3^{-1})\right] ^{d/2}} \prod_{j=1}^3 \left( \frac{V}{2 v_j} \right)^{\alpha_j-1}\exp\left[-\frac{1}{2}v_{j}\Omega_{j}^{2}\right]
\\
=\frac{\pi^{d/2}}{\left(2\right)^{\alpha_{t}-d/2}}\int_{0}^{\infty}\frac{1}{\Gamma^3}\frac{1}{\left(v_{1}^{-1}+v_{2}^{-1}+v_{3}^{-1}\right)^{d-\alpha_{t}}}\prod_{j=1}^{3}dv_{j}v_{j}^{\alpha_{t}-d/2-\alpha_{j}-1}\exp\left[-\frac{1}{2}v_{j}\Omega_{j}^{2}\right]
\end{eqnarray}
where $\Omega_{1}^2 = (\omega_2 - \omega_3)^2$, $\Omega_{2}^2 = (\omega_1 - \omega_3)^2$, and $\Omega_{3}^2 = (\omega_2 - \omega_1)^2$, and where we have used
\begin{equation}
V^{\alpha_{t}-d/2}=\frac{\left(v_{1}v_{2}v_{3}\right)^{\alpha_{t}-d/2}}{\left(v_{1}^{-1}+v_{2}^{-1}+v_{3}^{-1}\right)^{d/2-\alpha_{t}}}
\end{equation}
we also have used the shorthand $\Gamma^3=\Gamma[\alpha_1]\Gamma[\alpha_2]\Gamma[\alpha_3]$.

We now introduce a fourth Schwinger parameter for the sum $\sum v_i^{-1}$, which gives

\begin{equation}
=\frac{\pi^{d/2}}{\left(2\right)^{\alpha_{t}-d/2}}\int_{0}^{\infty}\frac{dtt^{d-\alpha_{t}-1}}{\Gamma\left[\alpha_{1}\right]\Gamma\left[\alpha_{2}\right]\Gamma\left[\alpha_{3}\right]\Gamma\left[d-\alpha_{t}\right]}\prod_{j=1}^{3}dv_{j}v_{j}^{\alpha_{t}-d/2-\alpha_{j}-1}\exp\left[-\frac{1}{2}v_{j}\Omega_{j}^{2}-\frac{t}{v_{j}}\right]
\end{equation}

Notice that if $d=2 \alpha_t$ this is not necessary and the integral is elementary (this is a special case of the magic identities \cite{Drummond:2006rz}). Continuing on, we would like to write this in terms of the modified Bessel function

\begin{equation}
K_{\nu}\left(z\right)=\frac{1}{2}\left(\frac{z}{2}\right)^{\nu}\int_{0}^{\infty}e^{-u-\frac{z^{2}}{4u}}u^{-\nu-1}du
\end{equation}

To do this, we simply change variables to $\frac{1}{2}v_{j}\omega_{kl}^{2}=u_{j}$, $du_{j}=\frac{1}{2}dv_{j}\omega_{kl}^{2}$,
$v_{j}=\frac{2u_{j}}{\omega_{kl}^{2}}$.

Our integral becomes

\begin{equation}
=\frac{\pi^{d/2}}{\left(2\right)^{\alpha_{t}-d/2}}\int_{0}^{\infty}\frac{dtt^{d-\alpha_{t}-1}}{\Gamma\left[\alpha_{1}\right]\Gamma\left[\alpha_{2}\right]\Gamma\left[\alpha_{3}\right]\Gamma\left[d-\alpha_{t}\right]}\prod_{j=1}^{3}du_{j}\left(\frac{2}{\Omega_{j}^{2}}\right)^{\alpha_{t}-d/2-\alpha_{j}}u_{j}^{\alpha_{t}-d/2-\alpha_{j}-1}\exp\left[-u_{j}-\frac{t\Omega_{j}^{2}}{2u_{j}}\right]
\end{equation}

\begin{equation}
=\pi^{d/2}\left(2\right)^{\alpha_t-d}\int_{0}^{\infty}\frac{dt\left(\sqrt{t}\right)^{d/2-2}}{\Gamma\left[\alpha_{1}\right]\Gamma\left[\alpha_{2}\right]\Gamma\left[\alpha_{3}\right]\Gamma\left[d-\alpha_{t}\right]}\prod_{j=1}^{3}\Omega_{j}^{-\alpha_{t}+d/2+\alpha_{j}}K_{d/2+\alpha_{j}-\alpha_{t}}\left(\sqrt{2t} \Omega_j \right)
\end{equation}

It turns out that this expression is sufficient to extract the divergence structure that we are interested in.  To see the relation to secular resonances, we need only the parts that include the frequency.

\end{document}